\documentclass[fleqn, 11pt]{article}

\usepackage{bm}

\begin{document}

\title{Classical Limit of Demagnetization in a Field Gradient}
\author{Leon Heller\\Biophysics Group,  P-21, Los Alamos National Laboratory, \\ Los
Alamos, New Mexico, 87544}
\date{}
\maketitle
\newcommand{\norm}{(2 \pi \hbar)^3}

\begin{abstract}
We calculate the rate of decrease of the expectation value of the transverse
component of spin for  spin-1/2 particles in a magnetic field
with a spatial gradient, to determine the conditions under which a previous classical
description is valid. A density matrix treatment is required for two reasons. The
first arises because the particles initially are not in a pure state due to 
thermal motion.  The second reason is that each particle  interacts with the
magnetic field and the other
particles, with the latter taken to be via a 2-body central force.  The equations
for the 1-body Wigner distribution functions are written in a general manner,
and the places where quantum mechanical effects can play a role are identified. 
One that may not have been considered previously concerns the momentum
associated with the magnetic field gradient, which is proportional to the time
integral of the gradient. Its relative magnitude compared with the important momenta
in the problem is a significant parameter, and if their ratio is not small some
non-classical effects contribute to the solution. 

	Assuming the field gradient is sufficiently small, and a
number of other inequalities are satisfied involving the mean wavelength, range of
the force, and the mean separation between particles, we solve the integro- partial
differential equations for the Wigner functions to  second order in the
strength of the gradient. When the same reasoning is applied to a different problem
with no field gradient, but  having instead  a gradient to the  z-component of
polarization, the connection with the diffusion coefficient is established, and we
find agreement with the classical result for the rate of  decrease of the transverse
component of magnetization. The corresponding result for this rate in the absence of
collisions is much greater. 
 
	An approximate value for the $^3$He self diffusion coefficient based on a 
hard core potential is found to be 23.1 mm$^2$/sec at room temperature and 7
atmospheres. This falls between the results obtained in an NMR experiment having
these values of temperature and pressure, in which the spatial gradient of the field
is time dependent, and earlier experiments using other techniques, corrected for
temperature and pressure variation. 

\end{abstract}

\section{Introduction}

The problem of a system possessing magnetization in the presence of a magnetic field
having a spatial gradient to its strength was considered classically by Carr and
Purcell
\cite{Carr}, Torrey
\cite{Torrey}, and Abragam
\cite{Abragam}. The rate of precession of  the transverse component of a particle's
magnetic moment depends on where it is located in the field, but thermal motion and
collisions smear this out. This was taken into account in \cite{Carr} by
carrying out a random walk, and in
\cite{Torrey} and \cite{Abragam} by adding a diffusion term to the
equation for the time rate of change of the magnetization. It leads to a
reduction of its transverse component which is exponential in the
cube of the time. Stejskal and Tanner \cite{Stejskal} took the same approach but
generalised the problem by permitting the gradient to be time dependent. In the
approximate solution obtained there \cite{Stejskal} the reduction of the
magnetization has a more complicated time dependence. A classical treatment of NMR
experiments involving spatial variation of the magnetization, including the effect of
RF pulses, is given in Sodickson and Cory
\cite{Sodickson}.  We shall examine the diminution of transverse magnetization as a
quantum mechanics problem to determine under what conditions the
classical approach described above is justified.
 
	An {\it isolated} spin 1/2 particle is described by a wave function the spin part
of which can be written as a linear combination of two basis states, which can be
taken to be eigenstates of the $z$-component of spin with eigenvalues $\pm 1/2$. Any
given values of the relative magnitudes and phases of the coefficients of those two
states represents a particle whose spin is pointing in some definite direction in
3-dimensional space. In the presence of a constant magnetic field the spin rotates
coherently about the field.

	The theoretical description of the process of interest is more complex and proceeds
in two steps, precisely following the order in which the experimental conditions are
created. Focussing on a single particle, it is located in a heat
bath, which means there is a classical probability distribution for its momentum
\cite{Wigner}. In addition, the method of preparation (partially polarized) creates a
probability distribution for its spin direction
\cite{Wolfenstein}. At this first stage already the momentum and spin variables for
the particle must be described by a density matrix \cite{Wigner, Wolfenstein, Zeh}.
In the presence of a {\it constant} magnetic field the probability distribution of
the particle's spin rotates coherently about the field direction, just as for
the case of an isolated particle in a pure state. At this stage the 
central interaction of the particle with the other particles, which are in the same
heat bath, has no effect on the spin.

	In the second step a gradient to the strength of the field is turned on. The
momentum distribution of the particles becomes distorted away from thermal
equilibrium because a magnetic moment in the direction of the field experiences a
force in the direction of the gradient. Now the 2-body interaction does have an
effect. The
expectation value of the transverse component of the particle's spin is
determined by the off-diagonal elements of the density matrix in spin space, and we
shall see that the rate at which those elements decrease with time is affected by
the interplay between the two-body interaction and the field gradient.

	The effect of (anti)symmetrization on transport properties was studied by Lhuillier
and Lalo\"{e} \cite{Lhuillier}, and included initially in the present work. Those
papers were not concerned with the effects of an applied magnetic field; and there
appears to be a technical difference between their treatment of the effect of 2-body
scattering on the 1-body density matrix and that given in Snider \cite{Snider},
which we follow. This is discussed in the text.

	In the Theory section we formulate the problem in terms of the 1-body Wigner
distribution functions, and the coupled integro-partial differential equations that
they satisfy  are obtained as well as the expressions for the expectation value of
the particle's spin in terms of those functions. Some important parameters are
identified, including the momentum associated with the magnetic field gradient.
Provided the gradient is sufficiently small and a number of other inequalities are
satisfied, the equations are solved in Section 3 by treating the gradient of the
magnetic field as a perturbation, to second order in its strength. The successive
terms of the solution are gathered together in Section 4, where the result  is
compared with the classical solution. The approximations encountered along the way
are summarized in Section 5.

	A word about terminology. The bulk of the paper is concerned with the quantum
mechanical calculation of the expectation value of the spin of a particle. The
classical calculation with which it is compared speaks about magnetization, which is
the magnetic moment per unit volume. The two quantities are simply related:
multiplying the expectation value of the spin by the magnetic moment of a particle
and by the number of particles per unit volume gives the magnetization since all the
particles are identical.

\section{Theory}

	The Hamiltonian for the system is taken to be a sum of 1-body terms,
$H^{(1)}$, plus a sum of 2-body potentials $V$. $H^{(1)}$ contains the
kinetic energy plus the interaction energy of the particle's magnetic
moment with the applied magnetic field, $B_z$, taken to be in the
$z$-direction. For spin-1/2 particles this becomes
\begin{equation}
	H^{(1)} = \frac{p^2}{2M} - \frac{\gamma}{2} \hbar \sigma_z B_z
			\label{eqn:1bodyHam}
\end{equation}
where $M$ is the mass of the particles, $\gamma$ is the gyromagnetic ratio
and $\sigma_z$ is the Pauli spin matrix. With the gradient of the
magnetic field strength in the direction of the unit vector $\mathbf{u}$, $B_z$ can
be written
\begin{equation}
	B_z = B_0 + \mathbf{u} \cdot \mathbf{x} G(t).
		\label{eqn:magfield}
\end{equation}
$B_0$ is the constant part of the field and $G(t)$ represents the
magnitude of the gradient of the field. 

	We follow the development
in Snider \cite{Snider}, but with the inclusion of (anti)symmetrization. There the
density matrix for an N particle system is reduced by taking successive traces down
to the 1-body density matrix,
$\rho$. It satisfies the equation
\begin{equation}
	\frac{\partial \rho(1)}{\partial t} = -\frac{i}{\hbar}
\left(\left[H^{(1)}(1),\rho(1)\right] + \mathop{\mathrm{Tr}}_{(2)}
\left[V(1,2),\rho^{(2)}(1,2)\right] \right)
		\label{eqn:parrho/t}
\end{equation}
where the arguments of the operators refer to particles
number 1 and 2 and $\rho^{(2)}$ is the 2-body density matrix. (A superscript (1) is
omitted on the  1-body $\rho$.) It is indicated in the second term within
the parenthesis in  (\ref{eqn:parrho/t}) that the trace is to be taken over the
variables of particle 2. The hierarchy of equations is cut off by neglecting the
terms in the equation for $\partial\rho^{(2)}/\partial t$ which involve
$\rho^{(3)}$, arguing that in a dilute gas when particles 1 and 2 are interacting the
interaction with a third particle can be neglected \cite{Snider}. With $n$ being the
number of particles per unit volume and $a$ the range of the two-body potential,
this requires that $a\ll d=n^{-1/3}$. Normalising to the conditions found in the
experiment reported by Schmidt et al. \cite{Schmidt} involving a partially polarized
gas of $^3$He atoms, which was performed at room temperature and a pressure of 7
atmospheres,
\begin{equation}
	\frac{a}{d} = 0.13 \bigg(\frac{P}{7 atm}\frac{293^\circ}{T}\bigg)^{1/3}
\bigg(\frac{a}{2.4 \cdot 10^{-10} m}\bigg) 
\end{equation}
Although the ratio $a/d$ may be on the borderline
of being sufficiently small for that particular experiment \cite{Schmidt}, we shall
neglect the possibility of triple collisions, and will continue to take
$a= 2.4 \cdot 10^{-10}$ m  as an estimate of the range of the He-He potential when
examining the size of other relevant parameters.   

	We assume that {\it prior} to a collision
between particles 1 and 2 they are uncorrelated, and hence $\rho^{(2)}$ is just the
product of the two 1-body density matrices; this corresponds to the Boltzmann
property for the classical distribution functions. Using scattering theory the effect
of a collision can be obtained as
\begin{equation}
	\rho^{(2)}(1,2) = \Omega^{(+)}\rho(1)\rho(2)\Omega^{(+)\dagger}
			\label{eqn:rho(1,2)}		
\end{equation}
where $\Omega^{(+)}$ is the wave operator \cite{Gell-Mann, Snider}. When this
expression for $\rho^{(2)}$ is inserted into
(\ref{eqn:parrho/t})
it becomes an equation for the 1-body density matrix $\rho$ alone. We
shall seek a solution of this equation to second order in the 
strength of the gradient $G(t)$. In Lhuillier and Lalo\"{e}
\cite{Lhuillier} instead of the wave matrix $\Omega^{(+)}$ in
Eq.(\ref{eqn:rho(1,2)}) the $S$-matrix is used. This relates the density matrix in
the distant past, before the collision, to its value in the distant future after the
collision is completed. But we are interested in the time development of $\rho$
during a collision \cite{Snider} and choose the wave matrix, therefore. Our
numerical results would be different had we chosen the $S$-matrix.

	 In the experiment reported by Schmidt et al. \cite{Schmidt}, a gas of $^3$He atoms
is partially polarized in the $z$-direction and fairly uniform spatially. At $t=0$ it
is tipped through an angle $\theta$ thereby producing a  component of spin in
the plane transverse to the direction of the magnetic field as well as a component
in the $z$-direction. It is the diminution of the transverse component with time that
was studied, and which we wish to evaluate. The most general form of the density
matrix for spin-1/2 particles can be written
\begin{eqnarray}
	\rho & = & \rho_I I + \bm{\rho} \cdot \bm{\sigma} \nonumber\\ 
	     & = & \rho_I I + \rho_z \sigma_z + \frac{1}{2} (\rho_+ \sigma_-
          +\rho_- \sigma_+)
		\label{eqn:rhospin}
\end{eqnarray}
where for any operator, $A_{\pm} = A_x \pm i A_y$, and the $\sigma_i$ are the Pauli
spin matrices. The four operators
$\rho_I$,  $\rho_z$, $\rho_+$, and $\rho_- = \rho_+^{\dagger}$  operate only in
ordinary (position or momentum) space. 

	It is convenient to work with the Wigner distribution function \cite{Wigner}
generalized to a 2x2 matrix to include the spin degree of freedom. This
generalization was also made in reference \cite{Lhuillier}       .  It is
defined by
\begin{eqnarray}
		\label{eqn:Wignerfnp}
	f_{mm'}(\mathbf{x},\mathbf{p},t) & = &
\frac{1}{(\hbar \pi)^3}
\int\frac{d^3q}{(2\pi \hbar)^3}
<\mathbf{p}-\mathbf{q},m |\rho| \mathbf{p}+\mathbf{q},m'>  
\exp(-2i\mathbf{x}\cdot\mathbf{q} / \hbar) \\
& = &																																																																				
		\label{eqn:Wignerfnx}		 
 \frac{1}{(\hbar \pi)^3} 
\int d^3 y<\mathbf{x}-\mathbf{y},m |\rho| \mathbf{x}+\mathbf{y},m'> 
\exp(2i\mathbf{p}\cdot\mathbf{y} / \hbar) 
\end{eqnarray}
where the indices $m$ and $m'$ take the values $\pm 1/2$. It follows from the
fact that $\rho $ is hermitean that the 2x2 matrix $f$ is also. We also need the
inverse relation
\begin{equation}
	<\mathbf{p},m|\rho|\mathbf{p'},m'> = \norm \int d^3x f_{mm'}(\mathbf{x,(p+p')}/2,t)
\exp[i(\mathbf{p'-p}) \cdot \mathbf{x}/ \hbar] .
		\label{eqn:rhomatel}
\end{equation}

 Taking the partial derivative of Eq.(\ref{eqn:Wignerfnp}) or
(\ref{eqn:Wignerfnx}) with respect to time and using Eq.(\ref{eqn:parrho/t})
gives
\begin{equation}
	\frac{\partial f_{mm'}}{\partial t} = S^K_{mm'} + S^{\gamma}_{mm'} +
S^V_{mm'}
		\label{eqn:partf/t} 
\end{equation}																																																																			
where $S^K$ arises from the kinetic energy term in $H^{(1)}$,  $S^{\gamma}$ from
the magnetic energy term in $H^{(1)}$, and $S^V$ from the 2-body potential
energy term.		$S^K$ is the well known flow term
\begin{equation}
		S^K_{mm'} = -\frac{1}{M} \mathbf p \cdot \bm \nabla f_{mm'}.
		 \label{eqn:S^K}
\end{equation}
	To obtain $S^\gamma$ requires evaluation of the commutator $[\sigma_z B_z,
\rho]$. Making use of Eqs. (\ref{eqn:magfield}) and (\ref{eqn:rhospin}) yields
\begin{eqnarray}
		\label{eqn:Sgam1/2}
	S^\gamma_{1/2} &  = & -\frac{\gamma}{2} \hbar G(t) \mathbf{u}
\cdot\bm{\nabla}_p f_{1/2}\\
		\label{eqn:Sgam-1/2}		 
	S^\gamma_{-1/2} & = & \frac{\gamma}{2} \hbar G(t) \mathbf{u}
\cdot\bm{\nabla}_p f_{-1/2}\\
		\label{eqn:Sgam+}
	S^\gamma_+ & = & -i\gamma B_z f_+
		\label{eqn:S^gam_+}	
\end{eqnarray}
where we have introduced abbreviated notations for all the   $S$ terms as
follows: $S_{1/2} \equiv  S_{1/2\ 1/2}$ ;  $S_{-1/2}
\equiv  S_{-1/2\ -1/2}$ ; and $S_+ \equiv  S_{-1/2\ 1/2}$; and the same
definitions apply to the components of the Wigner distribution functions $f$. We
shall be particularly interested in the function $f_+$, which involves the
off-diagonal elements of $\rho$ in the 2-dimensional spin space. In conjunction with
Eq.(\ref{eqn:partf/t}) for the time rate of change of the distribution functions, the
interpretation of Eqs. (\ref{eqn:Sgam1/2}) and (\ref{eqn:Sgam-1/2}) is as follows. A
magnetic moment having a component in the direction of (or opposite to) a magnetic
field whose strength has a spatial gradient will experience a force in the direction
of (or opposite to) that gradient. Eq.(\ref{eqn:Sgam+}) says that a magnetic moment
transverse to the direction of the magnetic field undergoes Larmor precession about
the field. 

	With the density matrix normalized to Tr($\rho)=N$, the total number of
particles, it is straightforward to obtain the following results for
$n$, the number of particles per unit volume, and $<\bm{\sigma
}>$, the expectation value of the particle's spin at position $\mathbf{x}$ and
time t
\begin{equation}
	n = \int d^3 p{}[f_{1/2} \mathbf{ (x,p)} + f_{-1/2} \mathbf{
(x,p)}]
		\label{eqn:n}
\end{equation}
and
\begin{eqnarray}
		\label{eqn:sigmasubz}	
	<\sigma_z> & = & \frac{\int d^3 p [f_{1/2} \mathbf{(x,p)}
- f_{-1/2} \mathbf{(x,p)}]}{\int d^3 p [f_{1/2} \mathbf{(x,p)}
+ f_{-1/2} \mathbf{(x,p)}]} \\
		\label{eqn:sigmasub+}
	<\sigma_+ > & = & 	 \frac{2 \int d^3 p	f_+ (\mathbf{x,p},t)}{\int d^3 p
[f_{1/2} \mathbf{(x,p)} + f_{-1/2} \mathbf{(x,p)}]} .
\end{eqnarray}
Once these quantities have been obtained the net magnetic moment per unit volume,
$\mathbf{M}$, can be obtained simply by multiplying $<\bm{\sigma}>$ by the magnetic
moment of a single particle, $\gamma \hbar /2$, and the number of particles per unit
volume, $n$.

	We now turn to $S^V$, the collision term in Eq.(\ref{eqn:partf/t}). 
Introducing the spin and momentum variables for particle 2, over which the trace
indicated in Eq.(\ref{eqn:parrho/t}) will be taken, and the antisymmetrization
operator
($I +\epsilon P_{12}$) where $P_{12}$ permutes all the variables of
particles 1 and 2 and $\epsilon =-1$ for fermions, $S^V$ can be written
\begin{eqnarray}
	\lefteqn{S^V_{mm'}\mathbf{(x,p)} = -\frac{i}{\hbar} \frac{1}{(\hbar \pi)^3}
\sum_{m_2} \int
\frac{d^3 q}{(2\pi \hbar)^3} \frac{d^3 p_2}{(2\pi \hbar)^3}} \nonumber \\ &
& \cdot <\mathbf{p-q}, m; \mathbf{p_2},m_2|(I+\epsilon  P_{12})[V,
\rho^{(2)}(1,2)]|\mathbf{p+q}, m';\mathbf{p}_2,m_2> \nonumber \\ & & \cdot
\exp(-2i \mathbf{x \cdot q}/\hbar).
		\label{eqn:S^V}
\end{eqnarray}
It is sufficient to antisymmetrize just one of the states because $V$
and $\rho^{(2)}$ are symmetric under the interchange of the two particles. 
In the Appendix we will develop the collision term  making
two approximations. The first is to assume that the 2-particle potential energy
$V$ consists purely of a central interaction. For example, at a separation of 2.4
\AA \ the moment-moment interaction energy of two $^3$He atoms is $\approx
10^{-8}$ mev, whereas the central interaction is $\approx 1$ mev. 
The second approximation amounts to assuming that the gradient of the magnetic field
strength is sufficiently small; this comes about as follows. 

Examination of 
Eqs.(\ref{eqn:partf/t}) through (\ref{eqn:Sgam+}), which represent
the contributions to the time derivative of the respective distribution functions,
shows that only
$S^\gamma _+$ contains dependence on position via the gradient term in the magnetic
field, $B_z$. Neglecting the collision terms $S^V$ for the moment, would lead to the
following conclusions. With the particle density $n$ and the expectation value of
the spin
 initially independent of position, $n$ and $<\sigma_z>$ remain so since
$f_{\pm 1/2}$ are independent of position.  $n$ is, of
course, also independent of time, and $<\sigma_z>$ as well since 
  $\sigma_z$ commutes with the Hamiltonian. On the other hand $f_+$  would contain a
factor 
$\exp[-i\gamma (B_0 t +\mathbf{u\cdot x}F(t))]$ as the only position dependence in
the problem, where 
\begin{equation} 
	F(t) = \int^t_0 dt' G(t') .
		\label{eqn:F(t)}
\end{equation} 
 We show below that all these conclusions remain valid with collisions
and
 write, therefore,
\begin{equation}
  f_+(\mathbf{x,p},t) = g_+(\mathbf{p})  \exp[-i\gamma (B_0 t +\mathbf{u\cdot
x}F(t))].  
		\label{eqn:f+form}
\end{equation}
The product $\gamma F(t)$ is the wave number associated with the magnetic field
gradient \cite{Sodickson}; in that reference it is designated as $k$.

	The consequences for 
the matrix elements of $\rho$ can be seen as follows. From Eq.(\ref{eqn:rhomatel})
those matrix elements with $m=m'$, since they  arise from $f_{\pm 1/2}$ which do
not depend on position, are local in momentum space,
\begin{equation}
	<\mathbf{p},m|\rho|\mathbf{p}',m> =	\norm f_m(\mathbf{p}) \norm \delta^{(3)}
(\mathbf{p'-p}).
		\label{eqn:rhomm}
\end{equation}
 On the other hand, for the off-diagonal matrix
elements we have
\begin{eqnarray}
	\lefteqn{
	<\mathbf{p},-1/2|\rho|\mathbf{p}',1/2> = \norm g_+((\mathbf{p+p'})/2) \exp(-i\gamma
B_0t) } \nonumber \\ & &
\cdot \norm \delta^{(3)}(\mathbf{p'-p}-\gamma \hbar \mathbf{u} F(t)) .
		\label{eqn:rho-1/21/2}
\end{eqnarray}

	Position dependence of a distribution function  is sometimes neglected {\it in the
collision term} on the assumption that the 
quantity that is varying (the strength of the magnetic field in the present case)
changes little over the distance of a mean free path \cite{Lifshitz}. But the
 $\delta$-function in  Eq.(\ref{eqn:rho-1/21/2})
contains another relevant quantity, $\gamma \hbar \mathbf{u} F(t)$, that has nothing
to do with the mean free path. It is a momentum that arises from the oppositely
directed forces felt by particles with their moments in the direction of the field,
or opposite to it, due to the field gradient.  After inserting two complete sets of
2-body spin and momentum states into  Eq.(\ref{eqn:S^V}) matrix elements of the type
$<\mathbf{p},-1/2|\rho|\mathbf{p}',1/2>$
 appear linearly in the collision term $S^V_+$ and bilinearly (or not at all) in
$S^V_{\pm 1/2}$. When the
$\delta$-function from Eq.(\ref{eqn:rho-1/21/2}) is combined with the two
momentum conserving $\delta$-functions from the  $t$- and $\Omega^{(+)}$-matrix
elements [see (\ref{eqn:tmatel}) and (\ref{eqn:omega+})] it has the following
effects. In Eq.(\ref{eqn:S^V}) for $S^V_{\pm 1/2}$ the integration variable
$\mathbf{q}$ is required to be zero; but in $S^V_+$ it becomes $\gamma \hbar F(t)
\mathbf{u}/2$. Consequently $S^V_+$ acquires a factor
$\exp[-i\gamma(B_0t+\mathbf{u\cdot x} F(t))]$, and as a result every term in
Eq.(\ref{eqn:partf/t}) for $\partial f_+/
\partial t$ has that factor. This confirms the statements above that $f_{\pm 1/2}$
are independent of position and $f_+$ has the structure shown in
Eq.(\ref{eqn:f+form}). No assumption about the magnitude of $F(t)$ has been made at
this point.

	If the magnitude of $\gamma \hbar F(t)$ were not sufficiently small compared to the
important momenta in the problem there would be significant quantum mechanical
consequences  for the collision terms of the displaced argument of the
$\delta$-function in Eq.(\ref{eqn:rho-1/21/2}). The arguments of $g_+$ and
the arguments of the $t$-matrix elements would be shifted by
amounts of order $\gamma
\hbar F(t) \mathbf{u}$. 
We shall delay discussion of the magnitude of the effect on
the collision terms due to the altered  $t$-matrix elements, but note here that it
would even lead to the presence of non-energy-conserving matrix elements. 

	The effect
of the variation in $g_+$ can be seen directly. Although the distribution functions
are of the Fermi-Dirac type, we shall be interested in conditions of negligible
degeneracy, $n\overline{\lambda}^3\ll1$, where
$\overline{\lambda}$ is the mean wavelength of the particles.  In that case we can
use the  Maxwell-Boltzmann distribution,  and its
  momentum derivative is of order $(p/\overline{p}^2)$ times
the function itself, where $\overline{p}$ is the mean momentum at temperature $T$.
The fractional shift in the magnitude of $g_+$, therefore, is of order $\gamma \hbar
F(t)/\overline{p}$.
  We shall  estimate it by normalising 
to the experimental conditions found in reference
\cite{Schmidt}. The gyromagnetic ratio for $^3$He is 204 MHz/Tesla, and the maximum
value of $F(t)$ in that experiment, designated $F_M$, is 3.5 $\cdot 10^{-7}$
Tesla-sec/cm. Estimating $\overline p$ by the mean value of the momentum at
temperature $T$, we obtain
\begin{equation}
	\frac{\gamma \hbar F(t)}{\overline p} \approx 9.5 \cdot 10^{-8}
\left (\frac{\gamma}{\gamma_{^3\mathrm{He}}}\right) \left
(\frac{F(t)}{F_M}\right) \left (\frac{M_{^3\mathrm {He}}}{M}
\frac{293^\circ}{T}\right) ^{1/2}.
		\label{eqn:g_+shift} 
\end{equation} 
For any other set of experimental parameters it is only necessary to insert the
appropriate values of the gyromagnetic ratio $\gamma$, the temperature $T$, the
molecular mass
$M$, and the integrated strength of the gradient of the magnetic field, $F(t)$. Note
that a larger strength of the field gradient or a smaller temperature would increase
the ratio in Eq.(\ref{eqn:g_+shift}), but the variation with $T$ is slow.

	For the remainder of the paper we shall assume that the shift in the arguments of
$g_+$ and the $t$-matrix elements can be neglected, and this leads to the expressions
for the  collision terms  presented in 
Eqs.(\ref{eqn:S^V1/2})
and (\ref{eqn:S^V+}). Eqs.(\ref{eqn:partf/t}) then become the
following set of coupled nonlinear integro- partial differential equations.
\begin{equation}
		 \frac{\partial f_{1/2}(\mathbf{p})}{\partial t} = - \frac{\gamma}{2} \hbar G(t)
\mathbf{u}
\cdot \bm{\nabla}_p f_{1/2}(\mathbf{p}) + S^V_{1/2}(\mathbf{p})  
		\label{eqn:partf1/2/t}
\end{equation}
\begin{equation}
		 \frac{\partial f_{-1/2}(\mathbf{p})}{\partial t} =  \frac{\gamma}{2} \hbar G(t)
\mathbf{u}
\cdot \bm{\nabla}_p f_{-1/2}(\mathbf{p}) + S^V_{-1/2}(\mathbf{p})  
		\label{eqn:partf-1/2/t}
\end{equation}
and in the equation for $\partial f_+/\partial t$ the derivative of the exponential
with respect to time cancels the term $S^{\gamma}_+$ leaving
\begin{equation}
 \frac{\partial g_+(\mathbf{p})}{\partial t}	= i\frac{\gamma}{M}\mathbf{u \cdot p}
F(t) g_+(\mathbf{p}) + \overline{S}^V_+(\mathbf{p}).
		\label{eqn:partg+/t}
\end{equation}

\section{Perturbation Expansion}

Our approach for solving these equations, consistent with the approximation
described above, is to treat the strength of the field gradient, $G(t)$, as a
perturbation. Since it is multiplied by the gyromagnetic ratio $\gamma$ we shall
expand the distribution functions up to second order in $\gamma$,
\begin{equation}
	f_{\pm 1/2} = f^{(0)}_{\pm 1/2} + \gamma f^{(1)}_{\pm 1/2} +\gamma^2 
f^{(2)}_{\pm 1/2} +O(\gamma^3)
		\label{eqn:f_pm1/2}
\end{equation}
and
\begin{equation}
	g_+ = g^{(0)}_+ + \gamma g^{(1)}_+ +\gamma^2 
g^{(2)}_+ +O(\gamma^3).
		\label{eqn:g_+} 
\end{equation}
This procedure converts the collision terms $S^V$ from being bilinear in the
functions $f$ and $g$ to being linear in the successive approximations $f^{(i)}$ and
$g^{(i)}$.

\subsection*{Zeroth Order}
The terms of order $\gamma^0$ in Eqs.(\ref{eqn:partf1/2/t}) - (\ref{eqn:partg+/t})
represent the initial condition in which the probability of finding the particle
with a given momentum is given by a thermal distribution, and the expectation value
of its spin is specified; both quantities are initially independent of position. 
Those distribution functions can be written 
\begin{equation}
	f^{(0)}_{\pm 1/2}(\mathbf{p}) = n_{\pm 1/2} B^{(M)}(p)
		\label{eqn:f^0_pm1/2}
\end{equation} 
and
\begin{equation}
	g^{(0)}_+(\mathbf{p}) = n_+ B^{(M)}(p) ,
		\label{eqn:g^0_+}
\end{equation}
where
\begin{equation}
	B^{(M)}(p) = (\beta^{(M)} /\pi)^{3/2} \exp (-\beta^{(M)} p^2)
\label{eqn:MaxB}	
\end{equation}
is the the Maxwell-Boltzmann distribution for a particle of mass $M$, and
$\beta^{(M)} = (1/2MkT)$.   As expected  from conservation of energy
\cite{Lifshitz}, when these zeroth order distribution functions are inserted in the
collision integrals they can be shown to vanish, confirming that
Eqs.(\ref{eqn:partf1/2/t}) - (\ref{eqn:partg+/t}) are satisfied to this order. The
unitarity relation gets used here and in the subsequent development [see Appendix].
The significance of the normalization constants can be seen from Eqs.(\ref{eqn:n}) to
(\ref{eqn:sigmasub+}), since they determine the {\it initial} values of the number
density,
$n$, and the expectation value of the spin, $<\bm{\sigma}>$: $n=n_{1/2}+n_{-1/2},\
<\sigma_z> =(n_{1/2}-n_{-1/2})/n$, and
$<\sigma_+>= n_+/n$. 

\subsection*{First Order}
	At the next order in $\gamma$ the distribution functions are distorted in the
direction $\mathbf{u}$ of the gradient, so we write
\begin{equation}
	f^{(1)}_{\pm 1/2}(\mathbf{p}) = n_{\pm 1/2} B^{(M)}(p)	\mathbf{u \cdot p}
h^{(1)}_{\pm 1/2}
\label{eqn:f^1_+-}
\end{equation}
and
\begin{equation}
	g^{(1)}_+(\mathbf{p}) = n_+ B^{(M)}(p)	\mathbf{u \cdot p}
h^{(1)}_+.
\label{eqn:g^1_+}
\end{equation}
In principle the functions $h^{(1)}$ could depend on the magnitude of the momentum,
 and a systematic expansion in a complete set
of Sonine polynomials, for example
\cite{Lifshitz}, could be undertaken. We shall take the $h^{(1)}$ functions to
be independent of $p$, thereby limiting that expansion to just the lowest term. Note
that these first order terms in the distribution functions contribute nothing to
the number density and the expectation value of the spin, Eqs. (\ref{eqn:n}) -
(\ref{eqn:sigmasub+}), because their integrals over all momenta vanish. As expected
if
$<\sigma_z>\not= 0$  the particles acquire a non-zero expectation value of 
momentum, and it can be shown to be  proportional to the product of
$(n_{1/2}-n_{-1/2})/n$ and $\gamma \hbar \mathbf{u} F(t)$. The latter factor is the
same one we required to be small compared to thermal momenta in the discussion
above. 

	Now pick out of Eqs. (\ref{eqn:partf1/2/t}) - (\ref{eqn:partg+/t}) all the terms of
first order in $\gamma$. Since the functions $h^{(1)}$ depend only on the time, these
become coupled ordinary first order linear differential equations
\begin{equation}
	n_{1/2} B^{(M)}(p) \mathbf{u \cdot p} \frac {d h^{(1)}_{1/2}}{dt} 
= n_{1/2} \hbar \beta^{(M)} B^{(M)}(p) \mathbf{u \cdot p} G(t) +S^{V(1)}_{1/2}
\label{eqn:parth^1_1/2/t}
\end{equation}
\begin{equation}
	n_{-1/2} B^{(M)}(p) \mathbf{u \cdot p} \frac {d h^{(1)}_{-1/2}}{dt} 
= -n_{-1/2} \hbar \beta^{(M)} B^{(M)}(p) \mathbf{u \cdot p} G(t) +S^{V(1)}_{-1/2}
\label{eqn:parth^1_-1/2/t}
\end{equation}
and
\begin{equation}
	n_+ B^{(M)}(p) \mathbf{u \cdot p} \frac {dh^{(1)}_+}{dt} 
= i \frac{1}{M}n_+ B^{(M)}(p) \mathbf{u \cdot p} F(t) +\overline S^{V(1)}_+ .
\label{eqn:parth^1_+/t}
\end{equation}
The expressions for the first order collision terms $S^{V(1)}$ are given in the
Appendix, where the above equations are developed into
Eqs.(\ref{eqn:dh^1_1/2/dt}) to (\ref{eqn:dh^1_+/dt}).

	There are three quantities in those equations, $I_U, I_I,$ and $I_\pi$ that depend
upon the details of the 2-body potential energy; see Eq.(\ref{eqn:I_U,I_I,I_pi}). 
$I_U$, which is the average over a Maxwell-Boltzmann distribution of the cube of
the momentum multiplied by the transport cross section, enters into the {\it direct}
terms. $I_I$ and $I_\pi$, which involve the $t$-matrix elements themselves and not
just their absolute square, appear only in the {\it exchange} terms, i.e., those
proportional to $\epsilon$, which arise from (anti)symmetrization of the wave
functions. To study the relative importance of the latter it is necessary to specify
the 2-body potential energy function for the problem of interest and calculate the
$t$-matrix that it produces. For He the interatomic potential becomes strongly
repulsive at distances less than 2.4 \AA \ \cite{Feltgen}, so we shall use a hard
sphere potential as a guide; and there are two limiting
cases for which analytic approximations to the scattering amplitude can be
obtained:  
$ka/\hbar \gg 1$ and $ka/\hbar \ll 1$ where k is the relative momentum of the
colliding particles and
$a$ is the radius of the hard sphere potential. Numerical calculations of the quantum
mechanical scattering amplitude in the short wavelength limit \cite{DSchmidt} are in
qualitative and even semi-quantitative agreement 
with the classical scattering of a scalar wave on a rigid sphere, as presented
in Morse and Feshbach \cite{Morse}, and also by the scattering of an electromagnetic
wave by a perfectly conducting sphere, given in Jackson \cite{Jackson}. That
$t$-matrix takes the approximate form 
\begin{equation}
 t(k,\theta) \approx \frac{2 \pi \hbar^2 }{\mu} \frac{a}{2}
\Big ( i\frac {ka}{\hbar}  
\frac{(1+ \cos \theta )J_1(\frac {ka}{\hbar} \sin \theta)}{ \frac {ka}{\hbar}\sin
\theta} 
 +  \exp [-2i\frac{ka}{\hbar} \sin \frac {\theta} {2}\ ]\  \Big ) \ \  
(ka/\hbar \gg 1)
		\label{eqn:t(k,theta)}
\end{equation} 
where the first term arises from scattering by the shadow side and the second term
from the illuminated side; $\mu$ is the reduced mass. The pure imaginary shadow
scattering amplitude dominates in the small angle region $\theta
\stackrel{\textstyle <}{\sim}
\hbar/ka
$, and oscillates rapidly with decreasing amplitude for larger angles.

	 Normalizing once again to the conditions in reference \cite{Schmidt} gives
\begin{equation}
	\frac{\overline{k}a}{\hbar} =12.5 \Big(\frac{M}{M_{^3\mathrm {He}}}
\frac{T}{293^\circ}
\Big ) ^{1/2} \frac{a} {2.4  \mathrm{\AA}}
		\label{eqn:ka/hbar}
\end{equation}
where we have chosen 2.4 \AA \ as the hard sphere radius for a gas of 
$^3\mathrm{He}$ atoms. Although it would require a very low temperature to get into
the long wavelength limit, even a moderately low temperature, $\approx 10^{\circ}$,
would introduce significant corrections to the $t$-matrix as well as the first order
collision terms, $S^{V(1)}$. In particular, contributions to $S^{V(1)}$ arising from
antisymmetrization of the wave functions would not be negligible and this would
produce important modifications to the remainder of the paper. Keeping that in mind
we now estimate the magnitudes of $I_U, I_I$ and $I_\pi$ in
Eq.(\ref{eqn:I_U,I_I,I_pi}) using the short wavelength expression
(\ref{eqn:t(k,theta)}). It should be noted that 
 this limit, $\overline{\lambda} /a \ll1$, together with the low density assumption,
$na^3
\ll1$, imply $n\overline{\lambda}^3 \ll1$, which is the negligible degeneracy
assumption made earlier. Using Eq.(\ref{eqn:t(k,theta)}) we can also verify the
assumption made earlier that the shift in the momentum (or momentum transfer)
argument of the $t$-matrix element by the amount $\gamma \hbar F(t)$ has a
negligible effect for the conditions of reference \cite{Schmidt}.

	Although the shadow scattering and the scattering from the illuminated side each
contribute $\pi a^2$ to the {\it total} cross section, the factor ($1-\cos \theta$)
in Eq.(\ref{eqn:sigma(k)}) for the {\it transport} cross section  eliminates
the shadow contribution, hence $\sigma_U(k) = \pi a^2$; and from
Eq.(\ref{eqn:I_U,I_I,I_pi}) $I_U=\pi \overline{k^3} a^2$, where $\overline{k^3}$ is
the average of $k^3$ over a Maxwell-Boltzmann distribution for a particle
having mass $\mu$. It takes a considerable amount of algebra to show that when the
short wavelength expression Eq.(\ref{eqn:t(k,theta)}) for the scattering amplitude is
inserted into  Eqs.(\ref{eqn:dsigmaI/dOmega}), (\ref{eqn:sigma(k)}) and
(\ref{eqn:I_U,I_I,I_pi}), $I_I$ is smaller than
$I_U$ by $O[(\overline{k}a/\hbar)^{-2}]$. $I_{\pi}$ is even smaller,
being  
$O (\exp[-(\overline{k}a/\hbar)^{-2}])$. Therefore all the terms
proportional to $\epsilon$ in Eqs.(\ref{eqn:dh^1_1/2/dt})-(\ref{eqn:dh^1_+/dt}) can
be neglected. 

	In the short wavelength limit
Eqs.(\ref{eqn:dh^1_1/2/dt})-(\ref{eqn:dh^1_+/dt}) for the time derivative of
the first order functions $h^{(1)}$ become
\begin{equation}
	n_{1/2} \frac{dh^{(1)}_{1/2}}{dt } =n_{1/2} \hbar \beta^{(M)} G(t)	+\frac{4
\beta^{(M)}}{3M}  n_{1/2}n_{-1/2}[h^{(1)}_{-1/2} - h^{(1)}_{1/2}] I_U
\end{equation}
\begin{equation}
	n_{-1/2} \frac{dh^{(1)}_{-1/2}}{dt } = - n_{-1/2} \hbar \beta^{(M)}
G(t) -\frac{4 \beta^{(M)}}{3M} n_{1/2}n_{-1/2}[h^{(1)}_{-1/2} - h^{(1)}_{1/2}] I_U
\end{equation}
and
\begin{equation}
	\frac{dh^{(1)}_+}{dt} = i\frac{1}{M} F(t) + \frac{4
\beta^{(M)}}{3M}  [n_{1/2}h^{(1)}_{1/2} + n_{-1/2}h^{(1)}_{-1/2} -(n_{1/2} +
n_{-1/2}) h^{(1)}_+] I_U 	.
\end{equation}
Summing the first two equations gives
\begin{equation}
	n_{1/2}h^{(1)}_{1/2} + n_{-1/2}h^{(1)}_{-1/2} = (n_{1/2} - n_{-1/2})
\hbar \beta^{(M)} F(t)
		\label{eqn:comb}
\end{equation}      
where $F(t)$ is the time integral of the field gradient,  
Eq.(\ref{eqn:F(t)}). There is no need to solve for
$h^{(1)}_{1/2}$ and $h^{(1)}_{-1/2}$ separately because only the combination shown
in Eq.(\ref{eqn:comb}) is needed in the equation for $dh^{(1)}_+/dt$; and
furthermore the terms of first order in $\gamma$ do not contribute to the
magnetization. 

	The solutions for the real and imaginary parts of
$h^{(1)}_+$ are
\begin{equation}
	\mathrm{Re}h^{(1)}_+ = \frac{4 \beta^{(M)}}{3M} (n_{1/2}-n_{-1/2}) \hbar \beta^{(M)}
I_U
\exp(-\alpha t) \int^t_0 dt'F(t') \exp(\alpha t')
		\label{eqn:Reh^1_+}		
\end{equation}
and
\begin{equation}
	\mathrm{Im}h^{(1)}_+ =\frac{1}{M} \exp(-\alpha t) \int^t_0 dt'F(t') \exp(\alpha t')
		\label{eqn:Imh^1_+}
\end{equation}
where
\begin{equation}
	\alpha = \frac{4 \beta^{(M)}}{3M}nI_U .
		\label{eqn:alpha}
\end{equation}
$n=n_{1/2}+n_{-1/2}$ is the total number of particles per unit volume. 

	$\alpha$ can also be expressed in terms of the diffusion constant, $D$, for a
related problem. Suppose there is no magnetic field and no transverse component of
the magnetization, but there is a spatial gradient to the $z$-component of
magnetization. This reduces to a standard 2-component problem with equal masses
(spin up and spin down); and if the diffusion constant is calculated with the same
approximation as above, namely keeping only the leading term in an expansion in
Sonine polynomials, then \cite{Massey}
\begin{equation}
	\alpha = \frac{kT}{MD} .
 	\label{eqn:DifCon}
\end{equation}

	 The qualitative significance of $\alpha^{-1}$ can be seen from Eq.(\ref{eqn:alpha})
or its expression in terms of the diffusion constant: it is the mean free time
between collisions.   Normalising to the conditions in reference
\cite{Schmidt}
\begin{equation}
	\alpha = 3.5\cdot 10^{10} \mathrm{sec}^{-1}\bigg(\frac{P}{7 \mathrm{atm}}\bigg)
\bigg(\frac{M_{^3\mathrm {He}}}{M}\frac{293^{\circ}}{T}\bigg)^{1/2} \bigg(\frac{a}
{2.4\cdot 10^{-10} \mathrm{m}}\bigg)^2 .
		\label{eqn:alphavalue}
\end{equation}
Provided $G(t)$, and therefore $F(t)$, vary on a time scale much longer than
$\alpha^{-1}$, the integrals in Eqs.(\ref{eqn:Reh^1_+}) and (\ref{eqn:Imh^1_+}) are
well approximated for $t>>\alpha^{-1}$ by
\begin{equation}
	\exp(-\alpha t) \int^t_0 dt'F(t') \exp(\alpha t') \cong \frac{F(t)}{\alpha} .
\end{equation}
This leads to
\begin{equation}
	\mathrm{Re}h^{(1)}_+ = \frac{n_{1/2}-n_{-1/2}}{n} \hbar \beta^{(M)} F(t)
		\label{eqn:Reh^1_+result} 
\end{equation}
and
\begin{equation}
	\mathrm{Im}h^{(1)}_+ = \frac{1}{M \alpha} F(t)
  \label{eqn:Imh^1_+result}		
\end{equation}
For conditions similar to those of reference \cite{Schmidt} $\mathrm{Re}h^{(1)}_+$ is
much smaller than $\mathrm{Im}h^{(1)}_+$, their ratio being, at most, that of the
mean wavelength to the mean free path; this will be small if the previous
assumptions that $\overline{\lambda}\ll a$ and $a\ll d$ are both satisfied, and we
shall neglect Re$h^{(1)}_+$. It vanishes exactly if there is no $z$-component  to
the expectation value of the spin.

\subsection*{Second Order}

There is a restriction on the form of a distortion of the 
 distribution functions $f_{\pm 1/2}$ that is of
second order in $\gamma$ (and therefore also of second order in the direction vector
$\mathbf{u})$, which arises from the requirement that it not alter the values of
the particle number density, $n$, and the expectation value of the $z$-component of
spin,
$<\sigma_z>$, both of which are fixed by the terms of zero order. Since these
quantities are obtained from Eqs.(\ref{eqn:n}) and (\ref{eqn:sigmasubz}), the
integral over all momenta  of the distortions must vanish. The only allowed form is
\begin{equation}
	f^{(2)}_{\pm 1/2}(\mathbf{p}) = n_{\pm 1/2} B^{(M)}(p) [\mathbf{u \cdot u}
-2\beta^{(M)}
\mathbf{(u \cdot  p)}^2] h^{(2)}_{\pm 1/2} .
\end{equation}
Indeed, such a form arises naturally from a second application of the operator
$\mathbf{u \cdot \nabla}_p$ from Eq.(\ref{eqn:partf1/2/t}) on $f^{(0)}_{\pm 1/2}$.  
 For $g^{(2)}_+$ we write
\begin{equation}
	g^{(2)}_+ = n_+ B^{(M)}(p) (\mathbf{u \cdot p})^2 h^{(2)}_+(t) ,
		\label{eqn:g^2_+}	
\end{equation}
which arises similarly from a second application of $\mathbf{u \cdot p}$  from
Eq.(\ref{eqn:partg+/t}) on $g^{(0)}_+$ .

	Now pick out of Eqs.(\ref{eqn:partf1/2/t})-(\ref{eqn:partg+/t}) for the time
derivatives of the distribution functions the terms of second order in $\gamma$.
When the equations for 
$\partial f^{(2)}_{\pm 1/2}/\partial t$ are integrated over all $\mathbf{p}$  they
become 0=0. The equation for the time derivative of
$g^{(2)}_+$ becomes
\begin{equation}  
	n_+ B^{(M)}(p) (\mathbf{u \cdot p})^2 \frac{dh^{(2)}_+}{dt} = i\frac{1}{M}
n_+B^{(M)}(p)	(\mathbf{u \cdot p})^2 F(t) h^{(1)}_+  
+ \overline{S}^V_+(\mathbf{p})                
 \end{equation}
Integrating this equation over all $\mathbf{p}$ 	and changing the variables of
integration in the collision term to ($\mathbf{P,k}$) that term can be shown to
contribute nothing; hence
\begin{equation}
	h^{(2)}_+(t) =i\frac{1}{M} \int^t_0 dt F(t) h^{(1)}_+(t) .
\end{equation}
Inserting the expression for Im$h^{(1)}_+$ from Eq.(\ref{eqn:Imh^1_+result}) gives
\begin{equation}
	h^{(2)}_+(t) = -\frac{1}{M^2 \alpha}  \int^t_0 dt F^2(t) .
		\label{eqn:h^2_+}
\end{equation}

\section{Time Dependence of $<\bm{\sigma}>$}

	We now gather together all the contributions to $<\bm{\sigma}>$, the expectation
value of the particle's spin, up to
 second order in the strength of
the gradient of the magnetic field, by
inserting into equations (\ref{eqn:sigmasubz}) and (\ref{eqn:sigmasub+}) the results
from (\ref{eqn:f+form}), 
 (\ref{eqn:f_pm1/2})  
 -(\ref{eqn:MaxB}) , (\ref{eqn:g^2_+}) and (\ref{eqn:h^2_+}). 
\begin{equation}
	<\sigma_z>  = <\sigma_z>(t=0) 
\end{equation}
and
\begin{eqnarray}
 \lefteqn{  
	<\sigma_+> = <\sigma_+> (t=0)  \cdot \exp[-i\gamma (B_0t+\mathbf{u\cdot x}F(t))]}
\nonumber\\
& &
 \cdot \bigg[1 - \gamma ^2 D  \int^t_0 dt F^2(t)  + O(\gamma^4)\bigg] ,
		\label{sigma_+2ndorder}
\end{eqnarray}
where we have made use of the expression for the diffusion constant $D$ from
Eq.(\ref{eqn:DifCon}). Eq.(\ref{sigma_+2ndorder}) represents the rate of decrease
of the expectation value of the transverse component of the particle's spin to
second order in the strength of the gradient of the magnetic field.

	How does this solution for $<\bm{\sigma}>$ compare with the procedure followed
in  references \cite{Carr}-\cite{Stejskal}  mentioned at the beginning of the
paper? This consists of adding a classical diffusion term to the equation for the
time rate of change of the magnetization, i.e.,
\begin{equation}
	\frac{\partial \mathbf{M}}{\partial t} = \gamma (\mathbf{M \times B}) 
+D \nabla^2 \mathbf{M} .
		\label{eqn:partM/t}
\end{equation}
Recall that $\mathbf{M}$ is given by  $<\bm{\sigma}>$ multiplied  by the magnetic
moment of a particle, $\gamma \hbar/2$, and the number of particles per unit volume.
With
$\mathbf{B}$ given in Eq.(\ref{eqn:magfield}) and the initial value of
$\mathbf{M}$ independent of postion, the exact solution of Eq.(\ref{eqn:partM/t}) is
that $M_z$ remains at its initial value; and $M_+$ is given by
\begin{equation}
	M_+(\mathbf{x},t) = M_+(t=0)\exp[-i\gamma (B_0t+\mathbf{u\cdot
x}F(t))]\exp \bigg[-\gamma^2D\int_0^t F^2(t)dt \ \bigg] .
	\label{eqn:M_+class}
\end{equation}
The final exponential factor is found in references \cite{Stejskal} and
\cite{Schmidt}. If that exponential is expanded for small argument, to second order
in
$\gamma$ it gives  the corresponding term from
Eq.(\ref{sigma_+2ndorder}). 

	It was stated in the Introduction that the rate of decrease of the transverse
component of magnetization is affected by both  thermal motion and the interaction of
a particle with its environment. By turning off the 2-body potential in
Eq.(\ref{eqn:partg+/t}), thereby setting the collision terms
$S^V$ to zero, the rate due just to the random velocity distribution can
be found, and the exact result is
\begin{eqnarray}
		\lefteqn {		
		M_+(\mathbf{x},t) = M_+(t=0)\exp[-i\gamma (B_0t+\mathbf{u\cdot
x}F(t))] \exp \bigg[- \gamma^2 \frac{kT}{M}
\bigg(\int^t_0F(t)dt\bigg)^2\bigg]} \nonumber \\
& &
	\hspace{10cm} \mbox{(no collisions)}  
\label{eqn:M_+nocoll}
\end{eqnarray}
Using expression (\ref{eqn:DifCon}) for the diffusion constant shows that the ratio
of the arguments of the exponentials in Eqs.(\ref{eqn:M_+nocoll}) and
(\ref{eqn:M_+class}) is qualitatively given by $\alpha T$ where $\alpha^{-1}$
is the mean free time between collisions and $T$ is the duration of the
gradient. For the conditions of the experiment in reference \cite{Schmidt} $\alpha
T \approx 10^7$. It is not surprising that the transverse component of magnetization
decreases more rapidly without collisions because the particles sample a much larger
range of magnetic field strengths in a given time than they would if experiencing
collisions, and consequently their precession rates cover a larger range.

	For the conditions of the  experiment in reference \cite{Schmidt}
the quantity $\gamma^2 \int^T_0F^2(t)dt$  has the value 0.032 sec/mm$^2$, where $T$
is the time at which the gradient is turned off. Using the experimental values
\cite{DSchmidt} of the ratio $|M_+(T)/M_+(0)|$ of the magnitudes of the transverse
magnetization (at various points along the length of the cylinder)  after
the gradient is turned off to its initial value, and comparing with 
Eq.(\ref{eqn:M_+class}), the value of the
$^3$He self diffusion coefficient is found to be $D$=15.9 mm$^2$/sec \cite{DSchmidt}.
The discrepancy with the values shown in Fig.5a of reference \cite{Schmidt} is due
to a numerical error in that reference \cite{DSchmidt}. Although the statistical
error in the value of
$D$ is small, only 0.4 mm$^2$/sec, there is a much larger systematic error of almost
25 percent  because the cell was prepared initially at much higher than
room temperature \cite{DSchmidt}.

	We also evaluated
$D$ from its relation to $\alpha$, Eq.(\ref{eqn:DifCon}), using the estimate  given
in (\ref{eqn:alphavalue}) based on a hard sphere potential with radius $a$=2.4 \AA,
and find
$D$= 23.1 mm$^2$/sec.  This falls between the results obtained in previous
experiments \cite{Barbe} and \cite{Liner}, after correcting for pressure and
isotopic mass dependence, and that reported in \cite{Schmidt, DSchmidt}.
		
\section{Summary}

The system under consideration consists of spin-1/2 particles in a magnetic field
with a time-varying gradient to its strength, and with its initial 
magnetization, taken to be spatially uniform, having components both transverse to
and along the field direction.   We now summarize the line of reasoning that led to
the expression (\ref{sigma_+2ndorder}) for the time development of the expectation
value of the transverse component of spin,
$<\sigma_+>$,  
	starting with Eq.(\ref{eqn:parrho/t})  for the time derivative of $\rho$,
the 1-body density matrix. It is the off-diagonal elements of $\rho$ in the
2-dimensional spin space, or equivalently the Wigner distribution function $f_+$,
that determine $<\sigma_+>$.

	A number of parameters play a role in this development and certain inequalities
among them were assumed, in succession. As each of these relations arose we compared
the relevant ratios with their values in the experiment reported by Schmidt et al.
\cite{Schmidt}, expressed in a way that allows quick comparison with any other set
of experimental parameters. For a low density system of spin 1/2
particles, in the equation for the 2-body density matrix we neglected 3-body
collisions, and this requires that the mean spacing between particles,
$d$, be large compare to the range of the force, $a$.  We also assumed that
{\it prior} to a collision the particles are uncorrelated, so that the 2-body density
matrix is the product of the respective 1-body density matrices. Quantum scattering
theory is then used to find the effect of a collision.

	After expressing the 1-body density matrix elements in terms of 
Wigner distribution functions, $f$ , which form a 2x2 matrix in spin space, a set of
coupled nonlinear integro- partial differential equations (\ref{eqn:partf/t}) for the
$f$'s is obtained. On the right hand side are three contributions to
their time rate of change: $S^K$ arising from the kinetic energy; $S^{\gamma}$ from
the magnetic energy; and
$S^V$ from the 2-body potential energy terms in the Hamiltonian. The first two terms
are simple to write down and are given in Eqs.(\ref{eqn:S^K}) -(\ref{eqn:S^gam_+}); 
the expressions for the collision terms
$S^V$ are obtained in the Appendix, Eqs.(\ref{eqn:S^V1/2}) and (\ref{eqn:S^V+}). Two
approximations were made to obtain those collision terms. The first is to assume
that the 2-body potential is purely central, and this is satisfied extremely well
for $^3$He atoms. The second assumes that the gradient of the magnetic field
strength is sufficiently small; this arises from the spatial dependence of the phase
factor that governs the Larmor precession, Eq.(\ref{eqn:f+form}), where $F(t)$ is
the time integral of field gradient. We expected and verified that this is the only
spatial dependence of the distribution functions. The product
$\gamma F(t)$ plays the role of a wave number, and leads to the off
diagonal element (in spin space) of the density matrix being non-local in
momentum space by the amount $\gamma \hbar F(t)$; see
Eq.(\ref{eqn:rho-1/21/2}). This nonlocality shifts the arguments of $g_+$ and the
$t$-matrix elements in the collision terms by amounts of the same order, and even
leads to non-energy-conserving matrix elements. If these shifts were sizable they
would represent important quantum mechanical effects. The magnitude of the effect
of the shift on
$g_+$ depends on the ratio of
$\gamma
\hbar F(t)$ to the mean thermal momenta, and in Eq.(\ref{eqn:g_+shift}) this is
shown to be very small for the conditions of the experiment in reference
\cite{Schmidt}. Using a hard sphere potential as a guide, this same small ratio
makes the effect of the shift in the arguments of the
$t$-matrix elements negligible as well.

	To solve the resulting equations (\ref{eqn:partf1/2/t}) - (\ref{eqn:partg+/t}) we
expanded the distribution functions in a perturbation series to second order in the
strength of the field gradient, $G(t)$, using the gyromagnetic ratio $\gamma$ as the
expansion coefficient since it multiplies $G$; this makes the equations linear in
each order.
 The only
distortions away from Maxwell-Boltzmann distributions that were used are of the form
$\mathbf{u
\cdot p}$ in first order and $\mathbf{u \cdot u}$ and $(\mathbf{u \cdot p})^2$ in
second order, where
$\mathbf{u}$ is the direction of the gradient. Omitting additional dependence on
$p^2$ is equivalent to keeping only the leading term in an expansion in Sonine
polynomials
\cite{Lifshitz}. 

	To first order in $\gamma$ the equations (\ref{eqn:dh^1_1/2/dt}) -
(\ref{eqn:dh^1_+/dt}) were written still containing the effect of antisymmetization
of the wave functions on the collision terms; these are the terms containing
$\epsilon$ and they involve, through the quantities $I_I$ and $I_{\pi}$ defined in
(\ref{eqn:I_U,I_I,I_pi}), not just the absolute square of the $t$-matrix elements but
also their real and imaginary parts.  For a hard sphere potential in the short
wavelength limit, however, the exchange terms  are smaller than the direct terms by
$O(\overline{\lambda} /a)^2$, where $a$ is the hard sphere radius, and
we expect this to be true for a more general potential with $a$ the range of the
force. But note from Eq.(\ref{eqn:ka/hbar}) that for $^3\mathrm{He}$ at room
temperature the ratio $a/\overline{\lambda}$ is only 12.5 with $a$ = 2.4 \AA. At a
significantly lower temperature corrections due to the exchange terms would become
important; they were neglected in the remainder of the paper. The direct terms, on
the other hand, manifest themselves through the single quantity
$I_U$, which is the mean of the product of the cube of the momentum with the
transport cross section
$\sigma_U$, averaged over a Maxwell-Boltzmann distribution for a particle having the
reduced mass. 

	Provided the field gradient  $G(t)$ varies on a time scale much longer than
the mean free time between collisions, $\alpha^{-1}$, the solution for the first
order distortion of the transverse Wigner distribution function  is given in
Eqs.(\ref{eqn:Reh^1_+result}) and (\ref{eqn:Imh^1_+result}). $\alpha$ is simply
related to the self diffusion constant $D$ for $^3$He, which can be more readily
realized in a different experiment with no field gradient but rather a spatial
gradient to the $z$-component of magnetization; this relation is given in
Eq.(\ref{eqn:DifCon}). The solution for the second order contribution to the
transverse distribution function follows readily from the first order term, and is
given in Eq.(\ref{eqn:h^2_+}).

	If all the restrictions mentioned above are satisfied, then the expression for the
 rate of decrease of the transverse component of spin, to second order in
the strength of the gradient, is given in Eq.(\ref{sigma_+2ndorder}). Comparison
with the classical treatment described in the Introduction, the exact solution for
which is given in Eq.(\ref{eqn:M_+class}), shows agreement to second order in the
strength of the gradient. It was also shown that the rate
would be much greater if there were no collisions, because the particles would sample
a much larger range of magnetic field strengths in a given time and consequently
their precession rates would also.

	It is not possible to make a good comparison between the magnitude of the $^3$He
self diffusion coefficient estimated from a hard sphere potential using
Eqs.(\ref{eqn:DifCon}) and (\ref{eqn:alphavalue}), and the experimental result in
\cite{Schmidt} because the temperature varied over  a large range
\cite{DSchmidt}. The theoretical estimate of $D$= 23.1 mm$^2$/sec falls between the
values obtained in that experiment and that obtained in previous experiments
\cite{Barbe,Liner}.

\section{Acknowledgments}

I am indebted to David Schmidt for introducing me to this subject and for many
informative discussions, and for valuable suggestions regarding the manuscript. I
also want to thank Emil Motolla and Wojciech Zurek for helpful conversations.

\renewcommand\theequation{\thesection-\arabic{equation}}
\appendix\section{Appendix}
\setcounter{equation}{0}


Here we work out the expressions for the collision terms $S^V$. After substituting
Eq.(\ref{eqn:rho(1,2)}) for $\rho^{(2)}(1,2)$ in Eq.(\ref{eqn:S^V})
the elastic scattering
$t$-matrix is introduced via the relations \cite{Gell-Mann, Snider} $\Omega^{(+)}
=(I+G^{(+)}t)$ and $V\Omega^{(+)} =t$, where $G^{(+)}$ is the Green's function.
Since the potential conserves
the total momentum and the $t$-matrix elements depend only on the relative
momenta of the two particles, they can be written
\begin{eqnarray}
		\lefteqn{	< \mathbf{p}_A, m_A; \mathbf{p}_B, m_B|t|\mathbf{p'}_A, m'_A ;
\mathbf {p'}_B, m'_B > {} = \delta _{m_A m'_A} \delta _{m_B
m'_B}} \nonumber \\ & &  \cdot (2 \pi \hbar)^3 \delta^{(3)}
(\mathbf {p}_A + \mathbf {p}_B -(\mathbf {p'}_A + \mathbf {p'}_B)) \,
t(\mathbf{k}_{AB}, \mathbf {k'}_{AB})
		\label{eqn:tmatel}
\end{eqnarray} 
where $\mathbf{k}_{AB}=(\mathbf{p}_A -\mathbf{p}_B)/2$. The corresponding
equation for the matrix element of $\Omega^{(+)}$ is
\begin{eqnarray}
		\lefteqn{	< \mathbf{p}_A, m_A; \mathbf{p}_B, m_B|\Omega^{(+)}|\mathbf{p'}_A,
m'_A ;
\mathbf {p'}_B, m'_B > {} = \delta _{m_A m'_A} \delta _{m_B
m'_B}} \nonumber \\
 & &  \cdot (2 \pi \hbar)^3 \delta^{(3)}
(\mathbf {p}_A + \mathbf{p}_B -(\mathbf{p'}_A + \mathbf{p}'_B)) \Omega^{(+)}
(\mathbf{k}_{AB},\mathbf{k}'_{AB})
		\label{eqn:omega+}
\end{eqnarray}
where
\begin{equation}
\Omega^{(+)}
(\mathbf{k}_{AB},\mathbf{k}'_{AB})	=
 [(2 \pi \hbar)^3
\delta^{(3)}(\mathbf {p}_A - \mathbf{p}'_A ) +
t(\mathbf{k}_{AB}, \mathbf{k}'_{AB})/(E'_{AB} + i\tau -E_{AB})]
		\label{eqn:Ommatel}
\end{equation}
%
with $E_{AB}$ the sum of the 1-body energies of particles A and B. (It is
understood that the limit $\tau \rightarrow 0$ is to be taken.) Since the magnetic
energy term in
$H^{(1)}$ is the same in the primed and unprimed states only the kinetic energy
contributes to the denominator in Eq.(\ref{eqn:Ommatel}). Hence
$E_{AB}$ can be taken to be $k^2_{AB}/2\mu$ where $\mu$, the reduced mass, is
$M/2$.

%

	As described in the text we are assuming that the magnitude of $\gamma \hbar F(t)$
is sufficiently small that the shift in the argument of the
$\delta$-function in Eq.(\ref{eqn:rho-1/21/2}) can be neglected.	 Hence we shall
write
\begin{equation}
<\mathbf{p},m|\rho|\mathbf{p'},m'>=\norm\delta^{(3)}(\mathbf{p-p'})
\rho_{mm'}(\mathbf{p})
		\label{eqn:rholocal}
\end{equation}
where
\begin{equation}
\rho _{mm}(\mathbf{p}) = \norm f_m(\mathbf{p})
\end{equation}
		\label{eqn:rhodiag} 
and
\begin{equation}
\rho _{-1/2 \ 1/2}(\mathbf{p}) = \norm g
_+(\mathbf{p})
\end{equation}
		\label{eqn:rho+}
%
%
	
	When two sets of intermediate states are introduced into Eq.(\ref{eqn:S^V}) 	two of
the integrations can be carried out immediately by making use of
Eqs.(\ref{eqn:tmatel}), (\ref{eqn:omega+}), and (\ref{eqn:rholocal}), giving
\begin{eqnarray}
		\lefteqn{S^V_{mm'}(\mathbf{x,p}) = -\frac{i}{\hbar}\frac{1}{\norm}\sum_{m_2} 
\int \frac{d^3p_2}{\norm} \frac{d^3p'}{\norm}\frac{d^3p'_2}{\norm}} \nonumber \\
& &
\cdot \norm \delta^{(3)} (\mathbf{p+p_2-(p'+p'_2)}) \nonumber \\
& & 
\{ [t(\mathbf{k,k'}) \Omega^{(+)*} (\mathbf{k,k'}) - 
\Omega^{(+)} \mathbf{(k,k')} t^*(\mathbf{k,k'})] \ 
\rho_{mm'}(\mathbf{p}')\rho_{m_2m_2}(\mathbf{p}'_2) \nonumber \\
& &
+\epsilon [t(\mathbf{-k,k'}) \Omega^{(+)*} (\mathbf{k,k'}) -
\Omega^{(+)} (\mathbf{-k,k'}) t^*(\mathbf{k,k'})] \ \rho_{m_2m'}(\mathbf{p}')
\rho_{mm_2}(\mathbf{p}'_2) \}
\end{eqnarray}
After substituting for $\Omega^{(+)}$ from Eq.(\ref{eqn:Ommatel}), for the terms
linear in the $t$-matrix elements the integrations over the primed variables can be
carried out, and for the terms bilinear in $t$ the energy denominators can be
combined, giving
\begin{eqnarray}
	\lefteqn{S^V_{mm'}(\mathbf {x,p}) = \frac{-i}{\hbar} \bigg \{ \frac{1}{\norm} \sum
_{m_2}
\int
\frac{d^3p_2}{\norm} \Big ( [t(\mathbf{k,k}) -t^*(\mathbf{k,k})]
\rho_{mm'}(\mathbf{p})
\rho_{m_2m_2}(\mathbf{p_2})} \nonumber \\
& &
+\epsilon [t(\mathbf{-k,k})\rho_{m_2m'}(\mathbf{p}) \rho_{mm_2}(\mathbf{p_2})
- t^*(\mathbf{-k,k})\rho_{m_2m'}(\mathbf{p_2}) \rho_{mm_2}(\mathbf{p}) ] \Big ) 
\nonumber \\
& &
+\frac{1}{\norm} \sum_{m_2} \int \frac {d^3p_2}{\norm} \frac {d^3p'}{\norm}
\frac {d^3p'_2}{\norm} \nonumber \\
& &
\cdot \norm \delta^{(3)} (\mathbf{p+p_2-(p'+p'_2)}) 2 \pi i \delta(E'-E) \nonumber \\
& &
\cdot[|t(\mathbf{k,k'})|^2  \rho_{mm'}(\mathbf{p'})
\rho_{m_2m_2}(\mathbf{p'_2}) +\epsilon t(\mathbf{-k,k'}) t^*(\mathbf{k,k'})
\rho_{m_2m'}(\mathbf{p'}) \rho_{mm_2}(\mathbf{p'_2}) ] \bigg \}
		\label{eqn:S^Vlong}
\end{eqnarray}
This establishes the fact that only energy conserving matrix elements contribute to
the collision terms, so $t(\mathbf{k,k'})$ can be written as  $t(k,\theta)$. 

	Note that any integral of the form 
\begin{eqnarray}
	\lefteqn{
	I(\mathbf{p,p}_2) = \int \frac {d^3p'}{\norm} \frac {d^3p'_2}{\norm} } \nonumber \\
& &
\cdot 
\norm \delta^{(3)} (\mathbf{p+p_2-(p'+p'_2)}) 2 \pi \delta (E'-E) 
F(\mathbf{p, p_2, p', p'_2)}
\end{eqnarray}
can be rewritten by changing the variables of integration to ($\mathbf{P',k'}$),
where $\mathbf{P'=p'+p'_2}$ and $\mathbf{k'=(p'-p'_2)/2}$. After carrying
out the integrations over $\mathbf{P}'$ and the magnitude of $\mathbf{k'}$, using
$E'-E=(k'^2-k^2)/2 \mu$, $I$ becomes
\begin{equation}
	I(\mathbf{p,p}_2) = 2 \pi \int \frac{\mu k d\Omega '}{\norm}
F(\mathbf{p,p}_2,\Omega ')
\end{equation}
where  $\Omega'$ stands for two angles specifying the direction  of $\mathbf{k'}$.
The combination
$\mu k d\Omega '/\norm$ is the density of final states.

	Making use of the relation between the differential scattering cross section and
the $t$-matrix elements gives
\begin{equation}
	v \frac{d \sigma _U}{d \Omega '} =\frac{2 \pi}{\hbar}
|t(\mathbf{k,k'})|^2 \frac{\mu k}{\norm}
\end{equation}
\begin{equation}
	v \frac{d \sigma _A}{d \Omega '} =\frac{2 \pi}{\hbar}
\Big |[t(\mathbf{k,k'}) + \epsilon t(\mathbf{k,-k'})]/\sqrt{2} \Big |^2 \frac{\mu
k}{\norm}
\end{equation}
and
\begin{equation}
	v \frac{d \sigma _I}{d \Omega '} =\frac{2 \pi}{\hbar}
t(\mathbf{k,-k'})t^*(\mathbf{k,k'}) \frac{\mu k}{\norm}
\label{eqn:dsigmaI/dOmega}
\end{equation}
where $\sigma_U$ and $\sigma_A$ are, respectively, the unsymmetrized and
antisymmetrized cross sections. $\sigma_I$ is not actually a cross section, just a
convenient notation for the interference term; $v=k/\mu$ is the relative
velocity of the colliding particles.

	Further progress is made by making use of rotational invariance of
the $t$-matrix, and unitarity of the $S$-matrix, which takes the form
\begin{eqnarray} 
	\lefteqn {
	2i \ \mathrm{Im} t(\mathbf{k,k}) = -i \int \frac{d^3p'}{\norm}
\frac{d^3p'_2}{\norm} |t(\mathbf{k,k'})|^2 } \nonumber \\
& &
\cdot \norm \delta^{(3)}(\mathbf{p+p_2-(p'+p'_2)}) 2\pi \delta(E'-E) ,
		\label{eqn:unitarity}
\end{eqnarray}
and from the expressions above this becomes
\begin{equation}
2i \ \mathrm{Im} t(\mathbf{k,k}) = -i \hbar \int d\Omega'	v \frac{d \sigma_U}{d
\Omega'} .
\end{equation}
The corresponding equation for $t(\mathbf{k,-k})$ is
\begin{equation}
2i \ \mathrm{Im} t(\mathbf{k,-k}) = -i \hbar \int d\Omega'	v \frac{d \sigma_I}{d
\Omega'} .
\label{eqn:Imt(k,-k)}
\end{equation}
	 In terms of these quantities the separate cases
of $S^V$ from Eq.(\ref{eqn:S^Vlong}) can be obtained by carrying out the summation
over $m_2$ and rearranging some terms.
\begin{eqnarray}
	\lefteqn{S^V_{1/2} (\mathbf{p}) = \int d^3p_2 d\Omega '  \Big
\{[f_{1/2}(\mathbf{p}') f_{-1/2}(\mathbf{p}'_2) - f_{1/2}(\mathbf{p})
f_{-1/2}(\mathbf{p}_2)] v\frac{d \sigma _U}{d \Omega '} } \nonumber \\
& &
+ [f_{1/2}(\mathbf{p}') f_{1/2}(\mathbf{p}'_2) - f_{1/2}(\mathbf{p})
f_{1/2}(\mathbf{p}_2)] v\frac{d \sigma _A}{d \Omega '} +\epsilon g_+(\mathbf{p}')
g_+^*(\mathbf{p}'_2) v\frac{d \sigma _I}{d \Omega '} \Big \} \nonumber \\
& &
-\epsilon \frac{i}{\hbar } \int d^3p_2 [t(\mathbf{k,-k})g_+(\mathbf{p})
g_+^*(\mathbf{p_2}) - c.c.]
		\label{eqn:S^V1/2} 
\end{eqnarray}
The corresponding equation for $S^V_{-1/2}$ is obtained from (\ref{eqn:S^V1/2}) by
making the interchanges $f_{1/2} \leftrightarrow \ f_{-1/2}$ and $g_+
\leftrightarrow \ g^*_+$. Finally, the expression for $\overline{S}^V_+$ becomes
\begin{eqnarray}
   \lefteqn{ \overline{S}^V_+(\mathbf{p}) = \exp[i\gamma (B_0t+\mathbf{u \cdot
x}F(t))]\  S^V_+(\mathbf{x,p}) =
  \int d^3p_2 d\Omega ' }
\nonumber \\
 & &
\cdot \Big \{ \Big(g_+(\mathbf{p'})[f_{1/2}(\mathbf{p'_2}) + f_{-1/2}(\mathbf{p'_2})]
- g_+(\mathbf{p})[f_{1/2}(\mathbf{p_2}) + f_{-1/2}(\mathbf{p_2})] \Big)
v\frac {d \sigma _U}{d \Omega '} \nonumber \\
& &
+\epsilon [g_+(\mathbf{p'})f_{-1/2}(\mathbf{p'_2})+
f_{1/2}(\mathbf{p'}) g_+(\mathbf{p'_2})] v\frac{d \sigma _I}{d \Omega '} \Big \}
\nonumber \\
& &
-\epsilon \frac{i}{\hbar } \int d^3p_2 \Big (t(\mathbf{k,-k}) 
[g_+(\mathbf{p}) f_{-1/2}(\mathbf{p}_2) + f_{1/2}(\mathbf{p}) g_+(\mathbf{p}_2)]
\nonumber \\
& &
-t^*(\mathbf{k,-k})[g_+(\mathbf{p}_2) f_{-1/2}(\mathbf{p}) + f_{1/2}(\mathbf{p}_2)
g_+(\mathbf{p})] \Big)
		\label{eqn:S^V+}
\end{eqnarray}

	We now extract the collision terms of first order in $\gamma$, to be used in
Eqs.(\ref{eqn:parth^1_1/2/t}) -  (\ref{eqn:parth^1_+/t}), by inserting
Eqs.(\ref{eqn:f^1_+-}) and (\ref{eqn:g^1_+}) in the above. It is helpful to
express each individual momentum variable in terms of the total and relative
momenta $\mathbf{P,k}$  and $\mathbf{k'}$. When this is done the terms in
$S^{V(1)}_{\pm 1/2}$ involving $d\sigma_A/d\Omega'$ vanish by conservation of
momentum; and making use of unitarity, Eq.(\ref{eqn:Imt(k,-k)}), the terms involving
the real part of $h^{(1)}_+$ cancel, leaving
\begin{eqnarray}
	\lefteqn{S^{V(1)}_{1/2}(\mathbf{p}) = \mathbf{u} \cdot \int d^3p_2 d\Omega'
B^{(M)}(p) B^{(M)}(p_2) \Big \{ n_{1/2} n_{-1/2} (\mathbf{p'-p})(h^{(1)}_{1/2}-
h^{(1)}_{-1/2})
v\frac{d\sigma_U}{d\Omega'}  } \nonumber \\
& &
+\epsilon |n_+|^2 2i\mathbf{k'} \mathrm{Im} h^{(1)}_+ v\frac {d\sigma_I}{d\Omega'} 
\Big \} \nonumber \\
& &
+\epsilon |n_+|^2 \frac{4}{\hbar} \mathrm{Im} h^{(1)}_+ \mathbf{u} \cdot \int d^3p_2
B^{(M)}(p) B^{(M)}(p_2) \mathbf{k} \mathrm{Re} t(\mathbf{k,-k}) .
\label{eqn:S^1_1/2}
\end{eqnarray}
Furthermore, $S^{V(1)}_{-1/2}(\mathbf{p})$ =  $- S^{V(1)}_{1/2}(\mathbf{p})$.
Following similar reasoning leads to
\begin{eqnarray}
	\lefteqn{\overline{S}^{V(1)}_+(\mathbf{p}) = \mathbf{u} \cdot n_+
\int d^3p_2 d\Omega'
B^{(M)}(p) B^{(M)}(p_2)  } \nonumber \\
& &
\cdot \Big \{ (\mathbf{p'-p}) [h^{(1)}_+ (n_{1/2} + n_{-1/2})-(n_{1/2}h^{(1)}_{1/2}+ 
n_{-1/2}h^{(1)}_{-1/2})] v \frac{d\sigma_U}{d\Omega'} \nonumber \\
& &
+ \epsilon \mathbf{k'} [(n_{1/2}h^{(1)}_{1/2}- n_{-1/2}h^{(1)}_{-1/2})
-(n_{1/2}-n_{-1/2})h^{(1)}_+]v\frac {d\sigma_I}{d\Omega'} \Big \} \nonumber \\
& &
-i\epsilon \frac{2}{\hbar}n_+ [(n_{1/2}h^{(1)}_{1/2}-n_{-1/2}h^{(1)}_{-1/2})
-(n_{1/2}- n_{-1/2})h^{(1)}_+] \nonumber \\
& &
\cdot \mathbf{u} \cdot \int d^3p_2 B^{(M)}(p) B^{(M)}(p_2)
\mathbf{k} \mathrm{Re}t(\mathbf{k,-k})
\end{eqnarray} 

	When these expressions for the first order collision terms are inserted into
Eqs.(\ref{eqn:parth^1_1/2/t}) - (\ref{eqn:parth^1_+/t}) we see that every term is of
the form
$\mathbf{u
\cdot}$  some vector, where
$\mathbf{u}$ is an arbitrary direction in space; therefore those equations must be
valid as vector equations with the $\mathbf{u}$ omitted. A necessary consequence can
be obtained by forming $\mathbf{p\cdot}$ those equations and integrating over all
$\mathbf{p}$. At the same time we make the following observations about the 
collision terms. Change the variables of integration from $(\mathbf{p,p_2})$ to
$(\mathbf{P,k})$, where $\mathbf{P}=\mathbf{p+p_2}$ and
$\mathbf{k}=(\mathbf{p-p_2})/2$, and make use of
\begin{equation}
	B^{(M)}(p) B^{(M)}(p_2) = B^{(2M)}(P) B^{(\mu)}(k)
\end{equation}
which follows from conservation of energy. For each
fixed value of
$\mathbf{k}$ choose its direction as the polar axis for the integration over the
angles of
$\mathbf{k'}$, so that $d \Omega'$ becomes $d \Omega = 2 \pi \sin \theta d\theta $
where $\theta$ is the angle between $\mathbf{k}$ and $\mathbf{k'}$. We also define
the following quantities.
\begin{eqnarray}
	\sigma_U(k) & = & \int _0 ^\pi d\Omega(1- \cos\theta) \frac{d\sigma _U }{d \Omega}
\nonumber \\
\sigma_I(k)  & = & \int _0 ^\pi d\Omega\cos\theta \frac{d\sigma _I }{d \Omega}
		\label{eqn:sigma(k)}
\end{eqnarray}
and
\begin{eqnarray}
I_U & = & \int d^3k k^3 B^{(\mu)}(k) \sigma_U(k) = <k^3 \sigma_U(k)> ^{(\mu)}
\nonumber \\  
I_I & = & \int d^3k k^3 B^{(\mu)}(k) \sigma_I(k) = <k^3 \sigma_I(k)> ^{(\mu)}
\nonumber \\
I_{\pi} & = & \int d^3k k^2 B^{(\mu)}(k) \mathrm{Re} t(k,\theta= \pi)
		\label{eqn:I_U,I_I,I_pi} 
\end{eqnarray}
From the definition of $d \sigma_I/d \Omega$, Eq.(\ref{eqn:dsigmaI/dOmega}), it is
seen that $\sigma_I(k)$ and $I_I$ are pure imaginary, whereas $I_U$ and
$I_{\pi}$ are real. $\sigma_U(k)$ is called the transport cross section
\cite{Lifshitz} and
$I_U$ is the mean value of $k^3\sigma_U(k)$ averaged over a Maxwell-Boltzmann
distribution for a particle of mass $\mu$.

	Eqs.(\ref{eqn:parth^1_1/2/t}) - (\ref{eqn:parth^1_+/t}) now become
\begin{eqnarray}
		\lefteqn{n_{1/2} \frac{dh^{(1)}_{1/2}}{dt } =n_{1/2} \hbar \beta^{(M)} G(t)} 
\nonumber \\
& &
+\frac{4 \beta^{(M)}}{3M} \Big \{n_{1/2}n_{-1/2}[h^{(1)}_{-1/2} - h^{(1)}_{1/2}] I_U
-2\epsilon |n_+|^2 \mathrm{Im}h^{(1)}_+  \mathrm{Im}I_I \Big \} \nonumber \\
& &
+ \epsilon \frac{8 \beta^{(M)}}{3 \hbar} |n_+|^2 \mathrm{Im}h^{(1)}_+ I_{\pi} ,
		\label{eqn:dh^1_1/2/dt}
\end{eqnarray} 
\begin{eqnarray}
		\lefteqn{n_{-1/2} \frac{dh^{(1)}_{-1/2}}{dt } = - n_{-1/2} \hbar \beta^{(M)}
G(t)} 
\nonumber \\
& &
-\frac{4 \beta^{(M)}}{3M} \Big \{n_{1/2}n_{-1/2}[h^{(1)}_{-1/2} - h^{(1)}_{1/2}] I_U
-2\epsilon |n_+|^2 \mathrm{Im}h^{(1)}_+  \mathrm{Im}I_I \Big \} \nonumber \\
& &
- \epsilon \frac{8 \beta^{(M)}}{3 \hbar} |n_+|^2 \mathrm{Im}h^{(1)}_+ I_{\pi}
		\label{eqn:dh^1_-1/2/dt}		
\end{eqnarray}
and
\begin{eqnarray}
		\lefteqn{\frac{dh^{(1)}_+}{dt} = i\frac{1}{M} F(t) + \frac{4
\beta^{(M)}}{3M} \Big \{ [n_{1/2}h^{(1)}_{1/2} + n_{-1/2}h^{(1)}_{-1/2} -(n_{1/2} +
n_{-1/2}) h^{(1)}_+] I_U }	 \nonumber \\
& &
+ \epsilon[(n_{1/2}h^{(1)}_{1/2} - n_{-1/2}h^{(1)}_{-1/2}) - (n_{1/2} -
n_{-1/2}) h^{(1)}_+] I_I \Big \} \nonumber \\
& &
-i \epsilon \frac{4\beta^{(M)}}{3 \hbar} [(n_{1/2}h^{(1)}_{1/2} -
n_{-1/2}h^{(1)}_{-1/2}) - (n_{1/2} - n_{-1/2}) h^{(1)}_+]I_{\pi} .
		\label{eqn:dh^1_+/dt}
\end{eqnarray}
These are four coupled ordinary linear differential equations for the four unknown
functions 
$h^{(1)}_{\pm 1/2}$ and the real and imaginary parts of $h^{(1)}_+$.

\end{document}